\documentstyle[aps,twocolumn]{revtex}
\begin{document}
\draft
\title{
Mesoscopic fluctuations of eigenfunctions and
``level velocity'' distribution in disordered metals
}
\author{Yan V. Fyodorov$^1\dagger$ and Alexander D. Mirlin$^2\dagger$}
\address{ $^1$ Fachbereich Physik, Universit\"at-GH Essen, 45117 Essen,
Germany}
\address{
$^2$ Institut f\"{u}r Theorie der Kondensierten Materie,
  Universit\"{a}t Karlsruhe, 76128 Karlsruhe, Germany}
\address{$\dagger$ Petersburg  Nuclear Physics Institute,
188350 Gatchina, St.Petersburg, Russia}
\date{July 8, 1994}
\maketitle
%\narrowtext
\begin{abstract}
We calculate
the distribution of eigenfunction amplitudes
and the variance of the ``inverse participation ratio'' (IPR) in
disordered metallic samples. A relation is established
between the distribution
function of IPR and that of ``level velocities'' --
parametric derivatives
of energy levels with respect to a random perturbation.

\end{abstract}
\narrowtext

\section{Introduction.}
Statistical properties of disordered metals
have attracted a considerable research interest last years.
It was understood that the old problem of a quantum particle
moving in a quenched random potential considered earlier in the
context of Anderson localization and mesoscopic phenomena\cite{meso}
exemplified a particular class of chaotic quantum systems and
had much in common with such paradigmatic problems in the domain of Quantum
Chaos as quantum billiards\cite{bill}.
The Wigner-Dyson energy level statistics first found in the
framework of random matrix theory (RMT) \cite{mehta} and considered to be a
``fingerprint'' of quantum chaotic systems \cite{bo} was shown to be relevant
for disordered metals as well\cite{efe,shkl}. This fact gave a boost to
a broad application of RMT results for qualitative and
quantitative description of mesoscopic conductors and stimulated a
common interest to statistical characteristics of spectra of disordered
systems\cite{Ak}.

At the same time less attention was paid to statistical properties
of eigenfunctions in disordered or chaotic
quantum systems. Recently, however, the distribution of eigenfunction
amplitudes was shown to be relevant for description of fluctuations
of tunnelling conductance across the ``quantum dots''\cite{Alh} as well as for
understanding some properties of atomic spectra\cite{gay}.
Besides, a so called ``microwave
cavity'' technique has emerged \cite{eksp} as a laboratory tool
to simulate a disordered quantum system. This
technique allows to observe
directly eigenfunctions spatial fluctuations and was used in
\cite{kud} to study experimentally the eigenfunction statistics
in weak localization regime.
All these facts make the issue of eigenfunction statistics
to be of special interest and are calling for a detailed theoretical
consideration.

In order to characterize eigenfunction statistics quantitatively, it
is convenient to introduce a set of moments $I_{q}=\int | \psi(r)|^{2q}
d^{d}r$ of eigenfunction local intensity $| \psi(r)|^{2}$\cite{ipr}. The
second moment $I_{2}$ is known as the inverse participation ratio (IPR).
This quantity is a useful measure of eigenfunction localization: it is
inversly proportional to a volume of a part of a system which
contributes effectively to eigenstate normalization. For completely
``ergodic'' eigenfunctions covering randomly, but uniformly the whole sample
$I_{2}\propto 1/V$, with $V$ being a system volume. If, in contrast,
the eigenfunctions are localized, i.e. concentrated in a region of
linear size $\xi$, the mean IPR scales as
$\overline{I_{2}}\propto \xi^{-D_{2}}$ where $D_{2}$ is an effective
dimension which can be different from a spatial dimensionality $d$ because
of a multifractal structure of eigenfunctions \cite{ipr}. Correspondingly,
IPR fluctuations reflect {\it level-to-level} variations of eigenfunction
spatial structure.

The most complete analytical study of statistical characteristics of
eigenfunctions was performed for the cases of $0d$ systems \cite{ep,epr},
as well as for strictly $1d$\cite{kol}
and quasi $1d$ \cite{fm,mf} geometry. Some analytical results were
obtained also for a system in the vicinity of localization
transition in the dimensionality
 $d=2+\epsilon,\,\,\,\epsilon \ll 1$ \cite{ipr}
as well as for $d\to \infty$\cite{mf2}. Let us note that in
Ref.\cite{ep,fm,mf,mf2,epr} the supersymmetry method
was used which is a very powerful tool to study distribution
functions of various quantities characterizing eigenfunctions statistics.

In the present article we address systematically the
issue of the eigenfunction
statistics for arbitrary spatial dimensionality $d$ in the weak
localization domain. In the leading approximation (which neglects
spatial structure of the system and treats it as a zero-dimensional
one) these statistics are
described by the RMT which predicts a Gaussian distribution of
eigenfunctions amplitudes $\psi(r)$ \cite{ep,mf}
for systems with unbroken or completely broken
time reversal symmetry \cite{Som}. It is known since
the paper by Altshuler and Shklovskii \cite{shkl} that the diffusion
motion of a particle in a metallic sample produces deviations of
spectral statistics from that predicted by RMT. To our best
knowledge, the analogous problem for the eigenfunctions statistics
for 2D or 3D systems has
never been studied. It is just considered in the present paper.
We use a recently developed method \cite{kr} which is based on the
supersymmetry technics \cite{efe,vwz} and combines a perturbative elimination
of fast diffusive modes (in spirit of renormalization group ideas) and
consequent non-perturbative evaluation of a resulting $0d$ integral.
In this way, we calculate the deviations
from the Gaussian distribution of $\psi(r)$
in mesoscopic metallic samples.
We calculate also the variance of the IPR which
turns out to be of order of $1/g^{2}$, where $g$ is the dimensionless
(measured in units of $e^2/h$)
conductance of the sample. In the final part of the paper we establish
a relation between the IPR distribution ${\cal P}_I(I_{2})$
and the distribution
${\cal P}_v(v)$ of the so-called ``level velocities'' $v$, which characterize
a motion of energy levels as a response to a small random perturbation.
It turns out that the finite value of IPR variance
 gives rise to deviations of the form
of ${\cal P}_v(v)$ from the Gaussian. The latter form is known to be
typical for completely ``ergodic'' systems
with unbroken or completely broken
time reversal invariance \cite{del,SA,Jap}.

\section{Distribution of eigenfunction amplitudes and IPR fluctuations.}
In order to calculate the distribution of eigenfunction amplitude and to
find the IPR variance we use the fact that relevant quantities
can be expressed
in terms of correlation functions of certain supermatrix
$\sigma$--model \cite{efe,vwz}.
A quite general exposition of the method can be found in \cite{FMRev}
 and is not repeated here.
Depending on whether the time reversal and spin rotation symmetries
are broken or not, one of three different $\sigma$--models is relevant, with
orthogonal, unitary or symplectic symmetry group. We consider first the case
of the unitary symmetry (broken time reversal invariance) and then
present the results for  two other cases.

\subsection{Unitary symmetry.}
The expressions for $\overline{I_{q}}$ and $\overline{I_{2}^{2}}$
(the bar standing for disorder averaging)
in terms of the $\sigma$--model read as follows \cite{fm,mf,FMRev}:
\begin{eqnarray}
 \overline{I_{q}}&=&
{-1\over 2V} \lim_{\epsilon\to 0}\epsilon^{q-1}{\left.
{\partial\over \partial u}\right|_{u=0}}\int DQ
\exp\{-{\cal F}^{(q)}(u,Q) \}\ ; \label{eq1} \\
\overline{I_{2}^{2}}&=&\frac{1}{6V}\lim_{\epsilon\to 0}
 \epsilon^{3}{\left.
{\partial^{2}\over \partial u^{2}}\right|_{u=0}}\int
DQ \exp\{-{\cal F}^{(2)}(u,Q)\}\ ;
\label{eq2}
\end{eqnarray}
where
\begin{eqnarray}
&& {\cal F}^{(q)}(u,Q)=\int d^{d}r\left\{
\frac{1}{t} \mbox{Str}
(\nabla Q)^{2} \right.\nonumber \\
&&\qquad\qquad\qquad \left.+ \epsilon\,  \mbox{Str}(\Lambda Q )+
u\, \mbox{Str}^{q} (Q\Lambda k)\right\}\ ; \nonumber \\
&& Q=T^{-1}\Lambda T\ ;\qquad
\Lambda=diag(1,1,-1,-1)\ ;\nonumber \\
&& k=diag(1,-1,1,-1)\ ;\qquad \frac{1}{t}=\pi\nu{\cal D}/4
\label{eq3}
\end{eqnarray}
Here $T$ is $4\times 4$ supermatrix belonging to the coset
space $U(1,1\mid 2)$, ${\cal D} $ is the classical diffusion constant, $\nu$
is the density of states and $V$ is the system volume.

Generally speaking, the RMT predictions are applicable to a disordered
metallic system under the following conditions: $L\gg l$; $E_{c}\gg \Delta$,
where $L$ is the system size, $l$ is  mean free path ,
$E_{c}={\hbar{\cal D} / L^{2}}$ is the Thouless energy and $\Delta$
is the mean level spacing.
Eigenfunctions for such systems are known to be ergodic with amplitudes
$\psi(r)$ being uncorrelated ( for $\mid r-r'\mid \gtrsim l$ ) Gaussian
distributed complex (real) variables for broken (unbroken)
time-reversal symmetry respectively. This immediately gives\cite{mf,epr}
 $\overline{I_{q}^{(u)}}={q!}/{V^{q-1}}$
and $\overline{I_{q}^{2}}=\overline{I_{q}}^{2}$,
where the superscript $u$ refers to the unitary symmetry.

In the framework of the  $\sigma-$model formalism these
results can be easily reproduced
if one neglects any spatial variation of the supermatrix field
$Q(r)$. Then eqs.(\ref{eq1}), (\ref{eq2}) are
reduced to integrals over a single supermatrix
which can be evaluted exactly. The corrections to RMT results have a form
of a regular expansion in small parameter ${\Delta/ E_{c}}=g^{-1}$.
The systematic way to construct such an expansion can be briefly outlined
as follows \cite{kr}. The matrix $Q(\bbox{r})$ is
decomposed as
\begin{equation}
Q(\bbox{r})=T_0^{-1}\tilde{Q}(\bbox{r})T_0\ ,
\label{adm1}
\end{equation}
where $T_0$ is a spatially uniform matrix and $\tilde{Q}$ describes all modes
with non-zero momenta.
When $\Delta\ll E_c$, the matrix $\tilde{Q}$ fluctuates only weakly
around the value
$\tilde{Q}=\Lambda$. It can be parametrized as \cite{efe}
\begin{eqnarray}
\tilde{Q}&=&\Lambda(1+W/2)(1-W/2)^{-1}\nonumber\\
&=&\Lambda\left(1+W+{W^2 \over 2}+{W^3\over 4}+{W^4\over 8}+\ldots\right)\ ,
\label{adm2}
\end{eqnarray}
where $W$ is a block off-diagonal supermatrix
representing independent fluctuating degrees of freedom.
Substituting this expansion into eqs.(\ref{eq1})--(\ref{eq3}) and integrating
out the ``fast'' modes  one obtains an
expression for renormalized functional ${\cal F}_{eff}^{(q)}(u,Q_0)$,
where $Q_0=T_0^{-1}\Lambda T_0$ is an $\bbox{r}$--independent matrix
(zero mode):
\begin{equation}
e^{{-\cal F}_{eff}^{(q)}(u,Q_0)}=\langle e^{{-\cal F}^{(q)}(u,Q_0,W)+\ln
J(W)}\rangle_W\ .
\label{adm3}
\end{equation}
Here $J(W)$ is the Jacobian of the transformation (\ref{adm1}),
(\ref{adm2}) from the variable $Q$ to $\{Q_0,\ W\}$; and
$\langle\ldots\rangle_W$ denotes integration over $W$. Now one expands
the exponential in (\ref{adm3}) and calculates	integrals over the fast
modes $W$ using the Wick theorem and the contraction rules
\cite{meso,efe}:
\begin{eqnarray}
&&\langle\mbox{Str} W(\bbox{r})PW(\bbox{r'})R\rangle \nonumber\\
&&\qquad=\Pi(\bbox{r},\bbox{r'})
(\mbox{Str}P\mbox{Str}R-\mbox{Str}P\Lambda\mbox{Str}R\Lambda)\ ;\nonumber\\
&&\langle\mbox{Str}[W(\bbox{r})P]\mbox{Str}[W(\bbox{r'})R]\rangle\nonumber\\
&&\qquad=\Pi(\bbox{r},\bbox{r'})
\mbox{Str}(PR-P\Lambda R\Lambda); \label{3a}
\end{eqnarray}
where $P$ and $R$ are arbitrary supermatrices.
Here $\Pi(\bbox{r}_1,\bbox{r}_2)$ is the diffuson propagator;
for an isolated sample	this propagator has the following form:
\begin{eqnarray}
&& \Pi(\bbox{r}_{1},\bbox{r}_{2})=\sum_{\bbox{q}}
\cos({\bbox{qr}_1})\cos({\bbox{qr}_2})\,\Pi(\bbox{q})\ ;\nonumber \\
&& \Pi(\bbox{q})
=\frac{1}{2\pi\nu V}\frac{1}{{\cal D}\bbox{q}^2+\epsilon}
\ ;\qquad  \bbox{q}=\pi\left(\frac{n_1}{L_1},
\ldots ,\frac{n_d}{L_d}\right)\nonumber\\
&& n_i=0,\pm 1,\pm 2,\ldots\ ;\qquad \sum n_i^2>0\ ;
\label{eq4}
\end{eqnarray}
where the system is thought to be of the size $L_1\times L_2\times...
\times L_d$. Finally, the integrals over $Q_{0}$ are performed
exactly, using the technique developed in \cite{efe,vwz}.

Applying this method to eqs.(\ref{eq1}), (\ref{eq3}) one obtains:
\begin{equation}
\overline{I_q^{(u)}}=\frac{q!}{V^{q-1}}\left\{1+\frac{a_1}{g}q(q-1)
+O\left(\frac{1}{g^2}\right)\right\}
\label{eq5}\end{equation}
where $g=2\pi\nu{\cal D}L^{d-2}$
is the conductance of the sample. The value of the coefficient $a_1$,
\begin{equation}
a_{1}=g\sum_{\bbox{q}}\Pi(\bbox{q})= {1\over \pi^2}
\sum_{n_i\ge 0;\ \protect\bbox{n}^2>0}
{1\over n_1^2+\ldots+n_d^2}\ ;
\label{eq5a}
\end{equation}
depends on the spatial dimension  and is equal to $a_{1}=1/6$ in quasi $1d$
systems. For $d \ge 2$ the corresponding  sum over momenta $\bbox{q}$ diverges
at large $|\bbox{q}|$ and is to be cut off at $|\bbox{q}|\sim l^{-1}$.
This gives $a_1=\frac{1}{2\pi}\ln{L/l}$ for $d=2$ and $a_1\propto L/l$ for
$d=3$.

Knowing all the moments $\overline{I_{q}^{(u)}}$ it is easy task to restore
the whole probability distribution ${\cal P}(y)$
of the eigenfunction local intensity  $y=V|\psi(\bbox{r})|^2$:
\begin{equation}
\label{eq6}
{\cal P}^{(u)}(y)=e^{-y}\left[1+\frac{a_1}{g}(2-4y+y^2)
+O\left({1\over g^2}\right)\right]
\end{equation}
The leading terms here reproduce the well-known RMT
result\cite{mehta}; the rest is
 the weak localization correction.

Now we turn to the consideration of IPR fluctuations.
It turns out that IPR variance is of order of $1/g^2$.	Thus the expression
(\ref{eq5}) is insufficient for our needs and should be extended
to the next order. By the same method we find:
\begin{eqnarray}
&& \overline{I_2^{(u)}}=\frac{2}{V}\left[1+\frac{2a_1}{g}+\frac{1}{g^2}(
2a_1^2-5a_2)+O\left(\frac{1}{g^3}\right)\right] \label{adm5} \\
&& \overline{\left[I_2^{(u)}\right]^2}=\left({2\over V}\right)^2
 \left[1+\frac{4a_1}{g} \right.\nonumber\\
&&\qquad\qquad \left.+\frac{1}{g^2}(8a_1^2-2a_2)+
O\left(\frac{1}{g^{3}}\right)\right] \label{adm6}
\end{eqnarray}
Here the coefficient $a_2$ is defined as
$a_2=g^2\sum_{\bbox{q}}\Pi^2(\bbox{q})$
and is equal to
\begin{equation}
 a_2={1\over \pi^4}
\sum_{n_i\ge 0;\ \protect\bbox{n}^2>0}
{1\over (n_1^2+\ldots+n_d^2)^2}\ ;
\label{a2}
\end{equation}
 the sum being convergent for $d<4$. In particular, for $d=1,2,3$ we
have $a_1=1/90\simeq 0.0111$, $a_2\simeq 0.0266$ and $a_3\simeq
0.0527$ respectively.

Thus, we find the following expression
for the relative variance of the IPR distribution:
\begin{equation}\label{eq9}
\delta^{(u)}(I_2)\equiv {\frac{\overline{\left[I_2^{(u)}\right]^2}
-\left[\overline{I_2^{(u)}}\right]^2}{\left[\overline{I_2^{(u)}}\right]^{2}}}
= \frac{8a_2}{g^2}+O\left(\frac{1}{g^3}\right)\end{equation}

\subsection{Orthogonal and symplectic symmetry.}

For the systems with unbroken time reversal symmetry (orthogonal and
symplectis $\sigma$-models) the calculation goes along the same lines.
The main difference is in the form of the contraction rules which in
this case include an additional (Cooperon) contribution
\cite{efe,IWZ,altland}
and can be presented in the following form:
\begin{eqnarray}
&&\langle {\mbox Str} W(\bbox{r}) P W(\bbox{r'}) R\rangle \nonumber\\
&&\qquad = 2\Pi(\bbox{r},\bbox{r'})
[\mbox{Str} P \mbox{Str} R-\mbox{Str} P\Lambda \mbox{Str} R\Lambda
\nonumber\\
 &&\qquad
+\alpha\mbox{Str}(P\Lambda\tilde{R}\Lambda-P\tilde{R})]\ ;
\nonumber \\
&&\langle \mbox{Str}[W(\bbox{r})P]
\mbox{Str}[W(\bbox{r'})R]\rangle \nonumber\\
&&\qquad =2\Pi(\bbox{r},\bbox{r'})
\mbox{Str}(PR-P\Lambda R\Lambda-P\tilde{R}+P\Lambda
\tilde{R}\Lambda)
\label{adm99}
\end{eqnarray}
where $\alpha=+1\ (-1)$ for the orthogonal (resp. symplectic)
symmetry, and
$\tilde{R}=M^{T}R^{T}M$ , with $M$ being a supermatrix representing
the time reversal transformation and
satisfying $M^{T}M=1$, $M^{2}=\alpha k$ (see \cite{altland}
for more details).

It turns out however that the formulae (\ref{eq5}), (\ref{eq9}) can be
generalized to the orthogonal and symplectic cases without additional
calculations. The leading weak--localization correction to
$\overline{I_q}$ originates from the one-diffuson-loop diagram which
is proportional to $q(q-1)a_1/g$. The numerical coefficient can be
restored by using the results we obtained earlier for the case of
quasi--1D systems \cite{mf,FMRev}:
\begin{eqnarray}
\overline{I_q^{(1D)}}&=&I_q^{(GE)}\{1+{1\over 6}q(q-1)x \nonumber\\
&+&{1\over
180}q(q-1)(q-2)(3q-1)x^2+O(x^3)\}\ ,
\label{adm4}
\end{eqnarray}
where $x=2/(\beta g)$; $\beta=1(2,4)$ for the orthogonal
(resp. unitary, symplectic) symmetry, and $I_q^{(GE)}$ denotes the
value of $I_q$ for the corresponding Gaussian ensemble. We see that
the quasi--1D result is in agreement with our general formulae
(\ref{eq5}), (\ref{adm5}). We restore therefore the coefficients of
the leading corrections to $\overline{I_q}$ in the orthogonal and
symplectic cases:
\begin{eqnarray}
\overline{I_q^{(o)}}&=&
\frac{(2q-1)!!}{V^{q-1}}\left\{1+\frac{2a_1}{g}q(q-1)
+O\left(\frac{1}{g^2}\right)\right\} \label{adm7}\\
\overline{I_q^{(sp)}}&=&
\frac{(q+1)!}{2^q V^{q-1}}\left\{1+\frac{a_1}{2g}q(q-1)
+O\left(\frac{1}{g^2}\right)\right\} \label{adm8}
\end{eqnarray}
The corresponding corrections to distributions of eigenfunctions
intensities are as follows \cite{note}:
\begin{eqnarray}
{\cal P}^{(o)}(y)&=&\frac{e^{-y/2}}{\sqrt{2\pi y}}
\left[1+\frac{a_1}{g}\left({3\over 2}-3y+{y^2\over 2}\right)
+O\left(\frac{1}{g^2}\right)\right]\label{adm8a} \\
{\cal P}^{(sp)}(y)&=&4y{e^{-2y}}
\left[1+\frac{a_1}{g}\left({3}-6y+{2y^2}\right)
+O\left(\frac{1}{g^2}\right)\right]
\label{adm8b}
\end{eqnarray}

As to the IPR variance, $\delta(I_2)$, the leading contribution to it
is given by the diagrams with a two-diffusons-loop and is therefore
proportional to $a_2/g^2$. The coefficient can be again restored by
comparison with the quasi--1D result \cite{mf,FMRev}:
\begin{equation}
\delta^{(1D)}(I_2)\simeq {16\over 45 \beta^2 g^2}
\label{adm9}
\end{equation}
In the quasi-1D case $a_2=1/90$, so that (\ref{adm9}) agrees with
(\ref{eq9}). For orthogonal and symplectic systems of a general
dimensionality, we find
\begin{eqnarray}
\delta^{(o)}(I_2)&\simeq&32a_2/g^2\ ; \label{adm10}\\
\delta^{(sp)}(I_2)&\simeq&2a_2/g^2 \label{adm11}
\end{eqnarray}
We have checked the above results by direct calculation in the
orthogonal case. It should be noted that in order to get the correct
results for $\overline{I_2}$ and $\overline{I_2^2}$ to $1/g^2$ order
in this case, one should take into account the Jacobian $J(W)$ in
eq.(\ref{adm3}) (whereas in the unitary case this Jacobian contributes
only starting from the $1/g^3$ order). For the orthogonal symmetry, we
find
\begin{equation}
\ln J(W) = -{1\over 4}\int \mbox{Str} W^2 + O(W^4)
\label{adm12}
\end{equation}
Carrying out the calculations we obtain
\begin{eqnarray}
&& \overline{I_2^{(o)}}=\frac{3}{V}\left[1+\frac{4a_1}{g}+\frac{1}{g^2}(
12a_1^2-30a_2)+O\left(\frac{1}{g^3}\right)\right] \label{adm13} \\
&& \overline{\left[I_2^{(o)}\right]^2}=\left({3\over V}\right)^2
 \left[1+\frac{8a_1}{g} \right.\nonumber\\
&&\qquad\qquad \left.+\frac{1}{g^2}(40a_1^2-28a_2)+
O\left(\frac{1}{g^{3}}\right)\right]\ ; \label{adm14}
\end{eqnarray}
and consequently, $\delta^{(o)}(I_2)=32a_2/g^2$, in agreement with
(\ref{adm7}), (\ref{adm10}).

\subsection{Discussion.}

Let us now give some comments on the results obtained in this section.
Eqs.(\ref{eq6}), (\ref{adm8a}), (\ref{adm8b}) present leading
(one--loop) corrections to the distribution of eigenfunction
amplitudes which are due to the weak--localization effects. It is easy
to see that the corrections lead to an increase in probability to find
a value of the amplitude  considerably smaller or considerably
larger than the average value. It is exactly what one would
intuitively expect assuming a tendency of eigenfunctions to
localization. It is interesting to note that the correction is of the
same type for all three ensemble, in contrast to the well known difference in
weak--localization corrections to the conductivity.

Deviation of the eigenfunctions amplitude distribution from its RMT
form was observed very recently in an experiment with a disordered
2D microwave cavity \cite{kud}. The form of the correction reported in
\cite{kud} is in agreement with our formula (\ref{adm8a}).

It should be noted that Eqs.(\ref{eq6}), (\ref{adm8a}), (\ref{adm8b})
are valid for not too large values of $y$: $y\lesssim \sqrt{g/a_1}$,
when the correction term is small as compared to the leading one. For
larger $y$ (i.e. in the far ``tail''), the distribution $\cal P(y)$
differs strongly from its RMT form and can not be found by the method
used in this paper. At present we are able to present the result for
the ``tail'' of ${\cal P}(y)$ for the quasi--1D geometry only. This can
be done using the exact analytical expressions for ${\cal P}(y)$ we have
found earlier for quasi--1d systems (see Eqs.(16), (17) of
Ref.\cite{mf} or Eqs.(79), (89) of Ref.\cite{FMRev}). We find three
different regions of the variable $y$:
\begin{eqnarray}
&&{\cal P}^{(u)}(y)\simeq e^{-y}\left[1+{x\over 6}(2-4y+y^2)+\ldots\right]\
;\ y\lesssim x^{-1/2} \label{adm15a}\\
&&{\cal P}^{(u)}(y)\sim \exp\left[-y+{1\over 12}y^2x+\ldots\right]\
;\ x^{-1/2}\lesssim y\lesssim x^{-1} \label{adm15b}\\
&&{\cal P}^{(u)}(y)\sim\exp\left[-4\sqrt{y/x}\right]\ ;\ y\gtrsim
x^{-1} \label{adm15c}
\end{eqnarray}
where $x=1/g$. For completeness, we repeated again the result of
preceding subsections, Eq.(\ref{adm15a}), which
 presents a relatively small correction to the RMT formula.
In the intermediate region (\ref{adm15b}), the correction {\it in the
exponent} is small compared to the leading term but much larger than
unity, so that ${\cal P}(y)\gg {\cal P}_{RMT}(y)$ though
$\ln{\cal P}(y)\simeq \ln{\cal P}_{RMT}(y)$. Finally, in the region
(\ref{adm15c}), the distribution function ${\cal P}(y)$ has nothing to
do with its RMT form. To understand the physical origin of the
asymptotical behavior (\ref{adm15c}) we note that it has exactly the
same form as in the case of a system of an {\it infinite} length, when
all states are localized (see Eq.(21) of Ref.\cite{mf} or Eq.(114) of
Ref.\cite{FMRev}). In the quasi--1D case $x=1/g=L/L_c$, where $L$ is the
sample length and $L_c$ is the localization length in an {\it infinite}
system. Thus, the ratio $y/x$ in the exponent in Eq.(\ref{adm15c}) is
equal to $y/x=V_c|\psi(\bbox{r})|^2$, where $V_c$ is the localization
volume. Therefore, the asymptotics (\ref{adm15c}) is controlled by the
correlation length $L_c$ and describes a probability to have a
quasi--localized ``bump'' with an extent much less than $L_c$
\cite{khmel}.
We stress again that the formulae (\ref{adm15b}), (\ref{adm15c}) have
been derived for quasi--1D systems; the problem of their
generalization to higher $d$ is still open.

The obtained value of the IPR fluctuations $\delta(I_2)\sim 1/g^2 \propto
L^{4-2d}$ is much larger than the RMT result $\delta(I_2)\propto 1/V \sim
L^{-d}$. These fluctuations have much in common with the universal
conductance fluctuations which have the same relative magnitude. It is
interesting to note that extrapolating this result to
 the Anderson transition point, where
$g\sim 1$, we get $\delta(I_2)\sim 1$, so that the magnitude of
fluctuations is of order of the mean value. Thus, despite the fact
that the wave function is extended through the whole sample, there is
no self--averaging of the IPR at the mobility edge.

\section{Distribution of level velocities.}
Let us now demonstrate that the IPR fluctuations  are intimately
related to level response characteristics of disordered system subject to
a random perturbation.

The issue of the energy level motion under an external perturbation
was under intensive study recently\cite{del,SA,cur,my}. In the paper\cite{SA}
the statistical properties of the ``level velocities''
 (LV) defined as $v_n={\partial E_n/ \partial \alpha}$, $E_n$ standing
for a given energy level and $\alpha$
 being some control parameter, were
studied for systems with completely ``ergodic'' eigenstates. In particular,
the LV distribution ${\cal P}(v)$ was shown to be Gaussian within the
zero-mode
approximation for supersymmetric $\sigma-$model.
On the other hand, in the recent paper \cite{my} the
LV distribution was calculated analytically for a long quasi $1d$ system
subject to a random perturbation. It was found that in the limit
$g\to 0$ which corresponds
to complete localization of eigenstates, ${\cal P}(v)$
is essentially nongaussian. The mean squared LV
$\langle\overline{v_n^2}\rangle$ was found
to be of order of inverse localization length $\xi^{-1}$, whereas
the RMT predicts $\langle\overline{v_n^2}\rangle\propto V^{-1}$.

In order to get understanding of these facts in broader context let us consider
a disordered system as described by a one-particle random
Hamiltonian ${\cal H}$. Let us impose on it some random perturbation ${\cal W}$
so that perturbed Hamiltonian is ${\cal H}+\alpha{\cal W}$, with
the parameter $\alpha$ controlling the strength of perturbation.
We assume that both the Hamiltonian ${\cal H}$ and the
perturbation ${\cal W}$ belong to the same pure symmetry class \cite{Jap}.
In the sequel, we denote the averaging over the realizations of
${\cal W}$ by $\langle\ldots\rangle$, which should be distinguished
from the averaging over unperturbed Hamiltonian ${\cal H}$ denoted
by bar.
The level velocity $v_{n}$ corresponding to an energy level $E_n$ can
be found within the conventional perturbation theory as
\begin{equation}\label{eq10}
v_{n}=\int d^d r\int d^d r' {\cal W}_{rr'}\psi_{n}^*(\bbox{r})
\psi_{n}(\bbox{r'})\ ,
\end{equation}
where ${\cal W}_{rr'}$ are matrix elements of the perturbation.
To be specific, these matrix elements are
supposed to be independent Gaussian distributed random variables with
the mean value equal to zero and the
variance $\langle {\cal W}^{*}_{rr'}{\cal W}_{rr'}\rangle={\cal W}_{0}
(|\bbox{r}-\bbox{r'}|)$. We also make a natural assumption
that the perturbation ${\cal W}$ is of finite range $\zeta$,
i.e. ${\cal W}_{0}(|\bbox{r}|\gg \zeta)$ is assumed to vanish.

By using eq.(\ref{eq10}) one easily obtains:
\begin{equation}\label{eq11}
\langle\overline{v_n^2}\rangle= \int d^d r\int d^d r' {\cal W}_{0}
(|\bbox{r}-\bbox{r'}|)
\overline{\mid \psi_n(\bbox{r})\psi_{n}(\bbox{r'})\mid^2}
\end{equation}
To simplify eq.(\ref{eq11}) we need to know the short-scale behaviour
of the correlator $\overline{|\psi_n^2(\bbox{r})\psi^2_n(\bbox{r'})|}$.
We have \cite{g4}:
\begin{equation}
\overline{|\psi^2(\bbox{r})\psi^2(\bbox{r'})|}=
\frac{1}{2}\left[1+k_d(|\bbox{r}-\bbox{r'}|)\right]
\overline{|\psi^4(\bbox{r})|}\ ,
\label{a}\end{equation}
where $k_d(\bbox{r})=\overline{\mbox{Im} G(\bbox{r})}^2/(\pi\nu)^2$ and
$G(|\bbox{r}-\bbox{r'}|)=\langle r|(E-{\cal H})^{-1}|r'\rangle$ is the Green
function of the Hamiltonian ${\cal H}$. The explicit form of $k_{d}(\bbox{r})$
for $d=2,3$ can be found in \cite{g4}; for our purposes it is enough
to mention its asymptotic behaviour:
\begin{eqnarray}
k_d(\bbox{r})\simeq 1\quad;\quad k_{f}r\ll 1\nonumber \\
k_d(\bbox{r})\ll 1 \quad;\quad k_{f}r\gg 1
\label{b}\end{eqnarray}
where $k_{f}$ is the Fermi momentum.

Substituting (\ref{a}) in (\ref{eq11}) and using (\ref{b}) we find
\begin{equation}
\langle \overline{v_n^2}\rangle = w_{0}I_2\ ;\qquad
 w_{0}=c\int d^d r {\cal W}_0
(\bbox{r})\ ,
\label{c}\end{equation}
where $I_2$ is the mean IPR introduced above and\\ $c=1\ (1/2)$ if
$k_f\zeta\ll 1$ (resp. $k_f\zeta\gg 1$).

Therefore, the mean squared LV $\langle\overline{v_n^2}\rangle$
is proportional to the mean IPR $\overline{I_2}$. In the infinite quasi $1d$
system $I_2$ is inversly proportional to the localization length
$\xi$ in agreement  with the behaviour
$\langle\overline{v_n^2}\rangle\propto \xi^{-1}$
found in \cite{my}.

This consideration can be extended to higher LV moments
$\langle\overline{v_n^{2q}}\rangle$
as well. After some manipulations we find the general
relation between the two distributions:
\begin{eqnarray}
{\cal P}_v(v)&\equiv& \langle\overline{\delta( v-v_n )}\rangle
\nonumber \\ &=& \int_{0}^{\infty}\frac{dI_2}{\left[2\pi
{w}_{0}I_2\right]^{1/2}}\exp{\left[-\frac{v^2}{2{w}_0 I_2}\right]}
{\cal P}_I(I_2)\label{eq12}
\end{eqnarray}

The following comment is appropriate here. For a given
realization of the random perturbation ${\cal W}$ one can define the mean
level velocity $\overline{v_n}$ (note that the averaging here goes only
 over the unperturbed Hamiltonuan ${\cal H}$, but not over ${\cal W}$).
 It follows from eq.(\ref{eq11}) that
$\overline{v_n}$ defined in such a way shows
Gaussian fluctuations around zero
from one realization of the perturbation to another.
The mean square $\langle \overline{v_n}^2\rangle$
can be easily found to be
$\langle \overline{v}^2\rangle=V^{-1} \int d^d r {\cal W}_0(\bbox{r})
|G(\bbox{r})|^2$  and is proportional to the inverse system volume $V^{-1}$.
At the same time the distribution function ${\cal P}_u(u)$
of the quantity $u=v-\overline{v_n}$ turns out to be self-averaging, i.e.
its form is the same for {\it any} realization of the perturbation ${\cal W}$.
The function ${\cal P}_u(u)$ can be
shown to satisfy the relation similar to eq.(\ref{eq12}):
\begin{equation}
{\cal P}_u(u) = \int_{0}^{\infty}\frac{dI_2\: {\cal P}_I(I_2)}{\left[2\pi
({w}_{0}I_2-\langle \overline{v}^2\rangle)\right]^{1/2}}
\exp{\left[-\frac{u^2}{2({w}_0 I_2-\langle \overline{v}^2\rangle)}
\right]}
\label{eq12a}
\end{equation}
The relation (\ref{eq12}) is then recovered as a convolution of
eq.(\ref{eq12a}) with the Gaussian distribution of $\overline{v}_n$.

 Eq.(\ref{eq12}) shows that the Gaussian form of
${\cal P}_v(v)$ found earlier for generic chaotic systems \cite{del,SA}
 is a consequence of the IPR being a selfaveraging quantity within zero-mode
approximation to nonlinear $\sigma-$model\cite{Jap}.
For disordered systems with
finite conductance $g$ the function ${\cal P}_v(v)$ becomes {\it nongaussian}
due to mesoscopic fluctuations of eigenfunctions. As a natural measure of the
deviations from the Gaussian form one can use the cumulant:
\begin{equation}\label{eq14}
\frac{\langle\overline{v_n^4}\rangle-
3\langle\overline{v_n^2}\rangle^2}{\langle\overline{v_n^2}\rangle^2}=
3\frac{\overline{I_2^2}-\overline{I_2}^2}{\overline{I_2}^2}
\equiv 3\delta(I_2)\propto {1\over g^2}
\end{equation}

With increasing disorder the conductance decreases and the system passes
to the strong localization regime $g\ll 1$. For this regime the explicit
 expression for both distribution
functions ${\cal P}_I(I_2)$ and ${\cal P}_v(v)$ were found
in quasi $1d$ systems in the course of
independent calculations \cite{fm,FMRev} (see also \cite{kolev})
 and \cite{my}. The results indeed satisfy the relation (\ref{eq12}).

 It could be interesting to study the
functions ${\cal P}_v(v)$
and ${\cal P}_I(I_2)$ also in $d=2$ and in the vicinity
of the Anderson transition for $d>2$ where their form is
expected to be universal.
Relation (\ref{eq12}) may provide
then a convenient background for investigations of
the statistical properties of eigenfunctions. In particular, the
behaviour of the mean square level velocity $\langle\overline{v^{2}}\rangle$
in the critical region can be used for
 extracting a value of the anomalous effective dimension $D_{2}$.

\section{Summary}
In this paper we have studied deviations  of the
eigenfunction statistical characteristics in a disordered metallic sample
from those predicted within the Random Matrix Theory. The found
correction to the distribution of eigenfunction amplitudes reflect the
influence of weak localization to an eigenfunction. The relative
magnitude of fluctuations of the inverse participation ratio is of the
order of $1/g^2$, where $g$ is the dimensionless conductance, in close
analogy with the conductance fluctuations.
 We have also  revealed the relationship between the
mesoscopic fluctuations of the inverse participation ratio and the form
of the ``level velocity'' distribution in a system.

\section*{Acknowledgments}
Y.V.F. is grateful to E.Akkermans and U.Sivan for stimulating his interest
in the issue of IPR fluctuations in metals
 and to S.Fishman, A.Kamenev and H.J.Sommers
for useful comments.
A.D.M. would like to thank V.E.Kravtsov for helpful comments.
The authors have much benefitted from discussions with
A.Kudrolli who communicated them results of  Ref.\cite{kud} prior to
 publication and are grateful to Y.Alhassid, Y.Gefen and F.Izrailev for
their interest to the work and numerous dicussions.
The hospitality of the Institute for Nuclear Theory at the University of
 Washington and the financial support from the
Alexander von Humboldt Foundation
 (A.D.M) and the program SFB237 ``Unordnung und Grosse Fluktuationen''
(Y.V.F.)  is acknowledged with thanks.


\begin{references}
\bibitem{note} For all symmetry cases, the conductance $g$ is defined
as $g=2\pi\nu{\cal D}L^{d-2}$, without including the factor $2$
due to the spin. In the symplectic case, the variable $y$ is defined
as the total (summed over the spin projection) local intensity:
$y=V(|\psi_\uparrow(\bbox{r})|^2+|\psi_\downarrow(\bbox{r})|^2)$.
\end{references}
\end{document}